\documentclass[a4paper]{article}

\usepackage{INTERSPEECH2020}
\usepackage{verbatim}
\usepackage{subcaption}

\title{Transformer with Bidirectional Decoder for Speech Recognition}
\name{Xi Chen$^1$, Songyang Zhang$^2$, Dandan Song$^3$, Peng Ouyang$^3$, Shouyi Yin$^{1,}$$^4$}
\address{
  $^1$Institute of Microelectronics, Beijing Innovation Center for Future Chip, Tsinghua University,
  $^2$ShanghaiTech University, $^3$Tsingmicro Intelligent Technology Co., Limited, \\
  $^4$Beijing National Research Center For Information Science And Technology}
\email{x-chen17@mails.tsinghua.edu.cn}

\begin{document}

\maketitle
\begin{abstract}
  Attention-based models have made tremendous progress on end-to-end automatic speech recognition(ASR) recently. However, the conventional transformer-based approaches usually generate the sequence results token by token from left to right, leaving the right-to-left contexts unexploited. In this work, we introduce a bidirectional speech transformer to utilize the different directional contexts simultaneously. Specifically, the outputs of our proposed transformer include a left-to-right target, and a right-to-left target. In inference stage, we use the introduced bidirectional beam search method, which can not only generate left-to-right candidates but also generate right-to-left candidates, and determine the best hypothesis by the score. 
  
  To demonstrate our proposed speech transformer with a bidirectional decoder(STBD), we conduct extensive experiments on the AISHELL-1 dataset. The results of experiments show that STBD achieves a 3.6\% relative CER reduction(CERR) over the unidirectional speech transformer baseline. Besides, the strongest model in this paper called STBD-Big can achieve 6.64\% CER on the test set, without language model rescoring and any extra data augmentation strategies.\footnote{This work was supported in part by the National Key R\&D Project (2018YFB2202600), the NSFC (61774094 and U19B2041), the China Major S\&T Project (2018ZX01031101002) and the Beijing S\&T Project (Z191100007519016). The authors thank Hao Ni and Long Zhou for their advices and helpful feedback.}
  
\end{abstract}

\section{Introduction}

End-to-end models for automatic speech recognition (ASR) have attracted much attention recently because of their concise pipeline, which integrate \textit{acoustic model}, \textit{pronunciation} and \textit{language model} into a single model. Comparing with conventional hybrid system, the various components can be optimized jointly in such an end-to-end framework. The most popular end-to-end speech recognition approaches typically include  \textit{connectionist temporal classification (CTC)}~\cite{graves2006connectionist, hannun2014deep,amodei2016deep,audhkhasi2018building}, \textit{recurrent neural network transducer (RNN-T)}~\cite{graves2012sequence, sainath2020streaming, li2020towards, he2019streaming}, and \textit{attention-based encoder-decoder architecture}~\cite{chan2016listen, pham2019independent, bahdanau2016end, dong2018speech, li2019improving}. 

The self-attention based methods (also called \textit{transformer})~\cite{vaswani2017attention}, have demonstrated promising results in various natural language processing (NLP) tasks recently. The transformer model exploits the short/long range context by connecting arbitrary pairs of position in the input sequence directly. Further more, the model can be trained in a parallel way, which is much more efficient than conventional recurrent neural networks. 

There have been several attempts to build an end-to-end system for speech recognition tasks based on the transformer model. Specifically, \cite{dong2018speech} first introduces the transformer model for the WSJ task. \cite{zhou2018syllable} explores different model units and finds the character-based model works best on the Mandarin ASR task. \cite{miao2020transformer, moritz2020streaming} demonstrates the transformer for streaming ASR. More recent efforts aim to enhance the transformer model on ASR by integrating connectionist temporal classification (CTC) for training and decoding~\cite{nakatani2019improving} or using unsupervised pre-training strategy~\cite{jiang2019improving}.

 However, these speech transformer models are typically trained in a left-to-right style, which means the decoding model predicts token depending on its left generated outputs, leaving the right-to-left contexts unexploited. Moreover, as the attention-based encoder-decoer network is an autoregressively generative model, they usually suffer from the issue of exposure bias in the sequence-to-sequence tasks. 
 
 Inspired by recent works~\cite{zhang2018asynchronous,zhou2019synchronous}, which introduce an interactive decoding model to exploit the bidirectional context for neural machine translation(NMT) task, we propose to utilize the right-to-left context and bidirectional targets on speech recognition tasks to tackle the aforementioned issues. 
 To this end, we propose a speech transformer model with bidirectional decoder structure. 
 The STBD consists of a shared encoder and a directional decoder which includes two different unidirectional decoders. The bidirectional model predicts the next left-to-right token and right-to-left token simultaneously, separately depending on the history contexts predicted by left-to-right and right-to-left decoding. Therefore, the outputs of our proposed transformer include a left-to-right target, and a right-to-left target. 
In inference stage, the bidirectional decoder generate both left-to-right and right-to-left targets with bidirectional beam search method. Finally, we keep the hypothesis with the highest score of different directional candidates. The extensive experiments and ablation study are conducted on the AISHELL-1 dataset~\cite{bu2017aishell}.
  
  The main contributions of this work can be summarized as follows:
  \begin{itemize}
  	\item We propose a novel speech transformer with bidirectional decoder to exploit the bidirectional context information for ASR task.
  	\item We investigate the effectiveness of the bidirectional target and bidirectional decoding with comprehensive experiments.
  	\item Our method achieves a 3.6\% relative CER reduction. The best model in this paper described as STBD-big achieves 6.64\% CER with a large margin improvement on AISHELL-1 dataset.
  \end{itemize}

\section{Method}
In this work, we introduce a novel speech transformer with bidirectional decoder for the ASR task. Below we start with a brief introduction of the conventional speech transformer used in previous works in Sec.\ref{sec:st}. We then present our network structure and beam search mechanism in Sec.\ref{sec:stbd}. Finally, we describe the training and inference process of our model in Sec.\ref{sec:train_infer}
\subsection{Speech Transformer}\label{sec:st}
The conventional speech transformer model~\cite{dong2018speech} has an encoder-decoder structure, the encoder maps an input sequence of acoustic feature 
$\mathbf{X}=\{\mathbf{x}_{1}, \ldots, \mathbf{x}_{n}\}$ 
to a sequence of hidden states 
$\mathbf{H}=\{\mathbf{h}_{1}, \ldots, \mathbf{h}_{n}\}$, 
Given $\mathbf{H}$, the decoder generates the outputs sequence
$\{y_{1}, \dots, y_{m}\}$ step by step, where $n$ is the frame number of input sequence, $m$ is the length of the output sequence. Both encoder and decoder are stacked with attention and position-wise feed-forward networks. 

  \vspace{-1em}
\subsubsection{Scaled Dot-product Attention}
The basic unit of the transformer model is self-attention mechanism, which is designed to map queries and a set of key-value pairs to outputs, where the queries $\mathbf{Q} \in \mathbb{R}^{t_{q} \times d_{q}}$, keys $\mathbf{K} \in \mathbb{R}^{t_{k} \times d_{k}}$, values $\mathbf{V} \in \mathbb{R}^{t_{v} \times d_{v}}$ and outputs are all vectors. The $t$ is the number of elements and $d$ is the feature dimensions of corresponding elements. We usually have $d_q=d_k, t_k=t_v$.

Firstly, queries compute dot-product with all keys, and then divide by scaled factor $\sqrt{d_{k}}$, which is used to prevent pushing the softmax function into extremely small regions. After softmax, we obtain attention scores which weight the values to generate final outputs. The process of scaled dot-product attention can be described as Eq~\ref{eq:dotattn}.
  \vspace{-1em}
\begin{equation}
  \text { Attention }(\mathbf{Q}, \mathbf{K}, \mathbf{V})=\operatorname{softmax}\left(\frac{\mathbf{Q} \mathbf{K}^{T}}{\sqrt{d_{k}}}\right) \mathbf{V}
  \label{eq:dotattn}
\end{equation}


\begin{figure*}[t]	
	\centering
	\begin{subfigure}[t]{0.45\linewidth}
		\centering
    \includegraphics[width=0.9\linewidth]{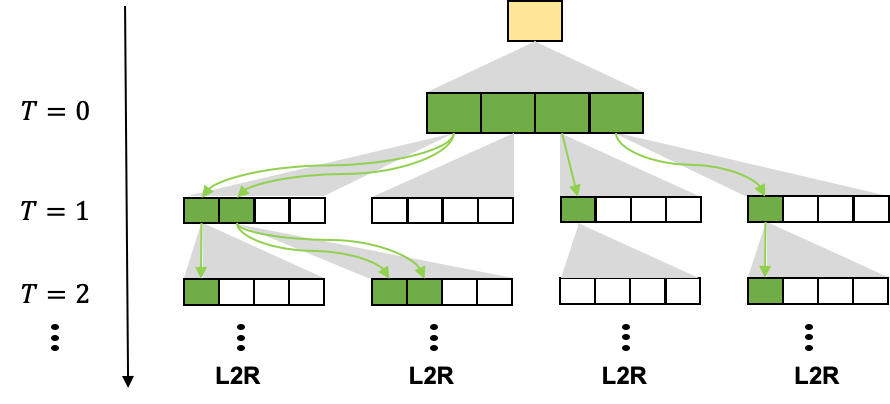}
    	\vspace{-1em}
		\caption{}\label{fig:1a}		
	\end{subfigure}
	\quad
	\begin{subfigure}[t]{0.45\linewidth}
		\centering
      \includegraphics[width=0.9\linewidth]{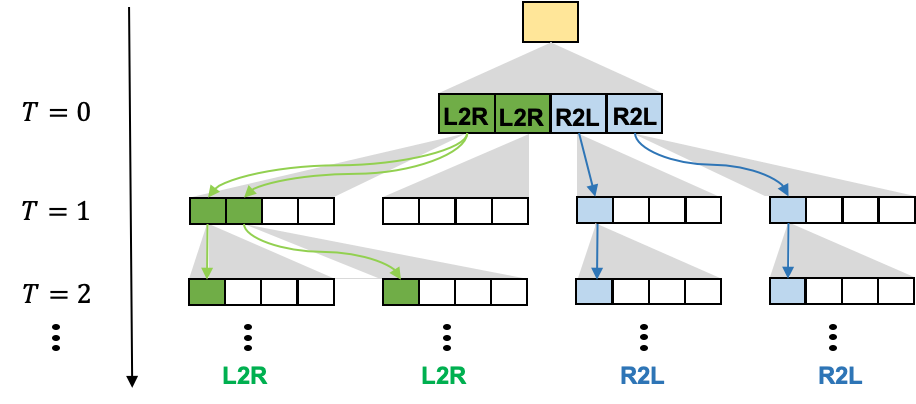}
      	\vspace{-1em}
		\caption{}
		\label{fig:1b}
	\end{subfigure}
	\vspace{-1em}
  \caption{a), Standard beam search with beam size 4, the green squares are alive beam. b) Bidirectional beam search with beam size 4.}
    \label{fig:sdbs}
\end{figure*}

 \vspace{-1em}
\subsubsection{Multi-head Attention}
In speech transformer, multi-head attention is applied to leverage different attending representations jointly. Specifically, it is beneficial to linearly project the queries, keys and values $h$ times with different, learned linear projections to $d_{k}$, $d_{k}$ and $d_v$ dimensions. Then the outputs of multi-head attention is generated by concatenating $h$ different head, which is illustrated in Eq~\ref{eq:multihead}.
  \vspace{-0.5em}
\begin{equation}
  \begin{aligned}
    & \text {MultiHead }(\mathbf{Q}, \mathbf{K}, \mathbf{V}) =\text {Concat}( \text{Head}_{1}, \cdots, \text {Head}_{\mathrm{h}}) \mathbf{W}^{O} \\
    & \text {Head}_{\mathrm{i}} =\text {Attention}\left(\mathbf{Q}\mathbf{W}_{i}^{Q}, \mathbf{K}\mathbf{W}_{i}^{K}, \mathbf{V}\mathbf{W}_{i}^{V}\right)
    \label{eq:multihead}
    \end{aligned}
\end{equation}
where  $\mathbf{W}_{i}^{Q}\in\mathbb{R}^{d_{m}\times d_{q}}$, $\mathbf{W}_{i}^{K}\in\mathbb{R}^{d_{m}\times d_{k}}$, $\mathbf{W}_{i}^{V} \in \mathbb{R}^{d_{m}\times d_{v}}$ are the parameters of linear projections, $\mathbf{W}^{O} \in \mathbb{R}^{h d_{v} \times d_{\text {m }}}$, $d_m$ is the feature dimension of the final outputs. Normally, $d_{q}=d_{k}=d_{v}=d_{\text {m}} / h$.
  \vspace{-1em}
\subsubsection{Position-wise Feed-Forward Networks}
In addition to attention sub-layers, each layer of encoder and decoder usually contains a fully connected feed-forward network, which consists of two linear layers with a ReLU activation between them. The outputs of feed-forward networks can be computed as Eq~\ref{eq:ff}.
  \vspace{-0.5em}
\begin{equation}
  \operatorname{F}(\mathbf{x})=\max \left(0, \mathbf{x} \mathbf{W}_{1}+\mathbf{b}_{1}\right) \mathbf{W}_{2}+\mathbf{b}_{2}
  \label{eq:ff}
\end{equation}
where $\mathbf{W}_1\in \mathbb{R}^{d_m\times d_f}$, $\mathbf{W}_2\in \mathbb{R}^{d_f\times d_m}$, $\mathbf{b}_1\in\mathbb{R}^{d_f},\mathbf{b}_2\in\mathbb{R}^{d_m}$, $d_f$ is the feature dimensionality of inner layer.

\subsubsection{Standard Beam Search}
With a trained speech transformer, beam search is used to find best hypothesis sentences for inference. At each step, we only remain the best $N$(called \textit{beam size}) hypothesis to generate the outputs of the next step. The process of standard beam search is illustrated in Fig.~\ref{fig:1a}.

\subsection{Speech Transformer with Bidirectional Decoder}\label{sec:stbd}
In order to exploit the right-to-left contexts, and improve the agreement between different directional hypothesis, we introduce the bidirectional decoder to speech transformer and denote our proposed model as \textit{STBD}.

\subsubsection{Structure of STBD}
\begin{figure}[t]
  \centering
  \includegraphics[width=0.85\linewidth]{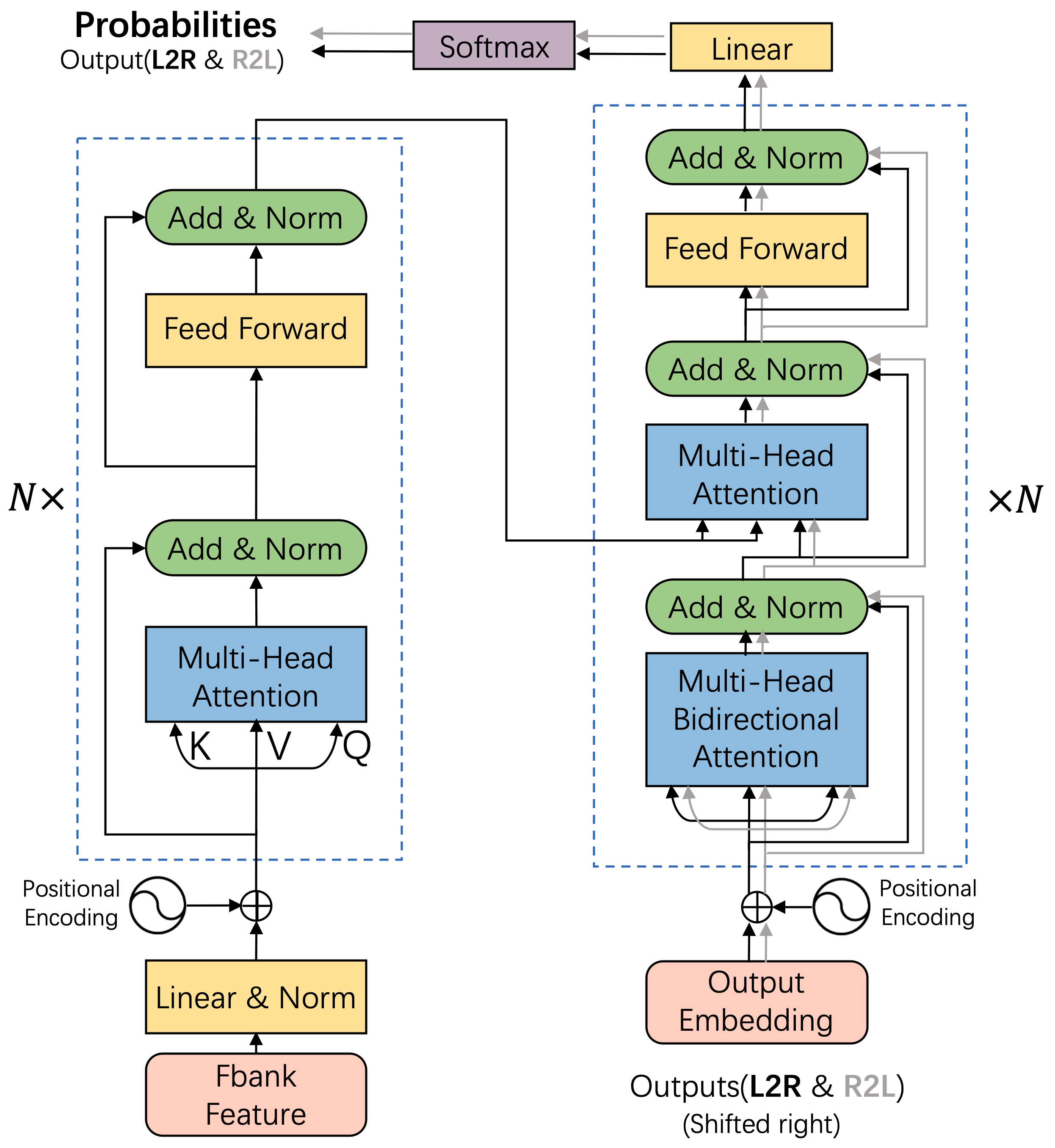}
  \vspace{-1em}
  \caption{The structure of bidirectional speech transformer}
  \label{fig:bd}
\end{figure}

Different with the conventional speech transformer, the STBD uses bidirectional targets to learn the encoder and decoder network.
The overall framework of speech transformer with bidirectional decoder(STBD) is illustrated in Fig.~\ref{fig:bd}. 

As the figure shows, the STBD includes an encoder and a bidirectional decoder, which consists of two unidirectional decoders with shared weight. The encoder is similar to standard encoder of speech transformer, which stacks $N$ times of sub-layers multi-head attention, feed-forward networks, and adds residual connection between the inputs and outputs of sub-layers. In STBD, there are two different unidirectional decoders, which decode from opposite direction and generate two different directional targets. 

\subsubsection{Bidirectional decoder}
In addition to left-to-right self-attention layers, the decoder contains right-to-left self-attention layers. The inputs of bidirectional dot-product attention consists of queries($[\overrightarrow{\mathbf{Q}}; \overleftarrow{\mathbf{Q}}])$, keys($[\overrightarrow{\mathbf{K}}; \overleftarrow{\mathbf{K}}]$), values($[\overrightarrow{\mathbf{V}}; \overleftarrow{\mathbf{V}}]$), which are concatenated by L2R and R2L states. The bidirectional hidden states $\overrightarrow{\mathbf{H}}$ and $\overleftarrow{\mathbf{H}}$ can be computed as:
\begin{equation}
  \overrightarrow{\mathbf{H}}=\text {Attention }(\overrightarrow{\mathbf{Q}}, \overrightarrow{\mathbf{K}}, \overrightarrow{\mathbf{V}})
\end{equation}
\begin{equation}
  \overleftarrow{\mathbf{H}}=\text {Attention }(\overleftarrow{\mathbf{Q}}, \overleftarrow{\mathbf{K}}, \overleftarrow{\mathbf{V}})
\end{equation}
The bidirectional attention also can be expanded into the multi-head form as:
  \vspace{-0.5em}
\begin{equation}
  \begin{aligned}
    & \text { MultiHead }([\overrightarrow{\mathbf{Q}} ; \overleftarrow{\mathbf{Q}}],[\overrightarrow{\mathbf{K}} ; \overleftarrow{\mathbf{K}}],[\overrightarrow{\mathbf{V}} ; \overleftarrow{\mathbf{V}}]) \\
    =& \text { Concat }\left(\left[\overrightarrow{\mathbf{H}}_{1} ; \overleftarrow{\mathbf{H}}_{1}\right], \ldots,\left[\overrightarrow{\mathbf{H}}_{h} ; \overleftarrow{\mathbf{H}}_{h}\right]\right) \mathbf{W}^{O}
    \end{aligned}
    \label{eq:mm}
\end{equation}

  The bidirectional decoder is expanded from the standard decoder of speech transformer. Specifically, it can be seen in Eq~\ref{eq:mm}, left-to-right decoder and the right-to-left decoder share the same network parameter.  
One unidirectional decoder decodes from left to right, generates the forward contexts, and the other one decodes from right to left, generates the backward contexts.
Instead of using $\left<\text{SOS}\right>$ as the start of sentences, we introduce two begin symbol $\left<\text{L2R}\right>$ and $\left<\text{R2L}\right>$, leading to generate different directional sentences separately.


\subsubsection{Bidirectional Beam Search}
Equipped with the bidirectional decoder, we also use the bidirectional beam search for sequence inference, which is illustrated in Fig.~\ref{fig:1b}. When the beam size is 4, we decode half of the beam from left to right, and decode the other half from right to left. At each time-step, the alive hypothesis sentences includes two candidates guided by $\left<\text{L2R}\right>$, and two candidates guided by $\left<\text{R2L}\right>$. The decoding process terminates when the $\left<\text{EOS}\right>$ symbol is predicted. Finally, the best hypothesis is determined by the scores of finished sentences. If the hypothesis sentence with the highest score is guided by $\left<\text{R2L}\right>$, we will reverse the final prediction.


\subsection{Training and Inference}\label{sec:train_infer}
In the training process, we jointly optimize the two different directional decoders with two directional targets, and the optimization objective is described as:
\vspace{-0.5em}
\begin{equation}
  \mathcal{L}_\text{total} = \alpha * \mathcal{L}_\text{L2R} + (1-\alpha)*\mathcal{L}_\text{R2L}
\end{equation}

\begin{table}[th!]
  \caption{CER Comparison with Previous Works on AISHELL-1 }
  \label{tab:cer}
  \centering
  \resizebox{0.5\textwidth}{!}{
  \begin{tabular}{ c c c c c}
    \toprule
    \textbf{Model} & 
    \textbf{Enc-Dec-Dm-Head} &
    \textbf{Dev CER} &
    \textbf{Test CER} &
    \textbf{Para}\\
    \midrule
    LFMMI\cite{bu2017aishell} & -&6.44\% & 7.62\% & -\\
    LAS\cite{shan2019component} & -  & -  & 10.56\%& - \\
    ST\cite{fan2019speaker} & 6-6-512-16&  - & 8.23\% & -\\
    SAST\cite{fan2019speaker} &6-6-512-16  & 7.59\% & 7.82\%& -\\
    SA-T+PA\cite{tian2019self}   &-&8.3\% & 9.30\%& -\\
    \midrule
    ST-L2R & 8-4-512-8 &  6.52\%  & 7.45\%  & 56.9M    \\
    ST-R2L & 8-4-512-8 &  6.62\%  & 7.45\%  & 56.9M    \\
    \midrule
   	\textbf{STBD}  & 8-4-512-8 &  \textbf{6.23\%} & \textbf{7.18\%} & 56.9M\\
    \textbf{STBD-Big} & 8-4-1024-16& \textbf{5.80\%} & \textbf{6.64\%} & 222.9M\\
    \bottomrule
  \end{tabular}
  }
\end{table}

The $\mathcal{L}_\text{L2R}$ and $\mathcal{L}_\text{R2L}$ are cross-entropy loss computed for two directions. The ratio factor $\alpha\in(0,1)$ characterizes the preference of the model. In this paper, we suppose that both directions are equally important, and set it to $\alpha = 0.5$.

\section{Experiment}
We evaluate our methods on the speech recognition by conducting a set of experiments on AISHEEL-1 dataset. In this section, we introduce the experimental setup and report detailed experimental results.
\subsection{Experimental Setups}
We demonstrate our method on the Chinese Mandarin dataset AISHELL-1~\cite{bu2017aishell}, which contains about 178 hours of Mandarin speech audio. The input data is a sequence of 80-dim fbank acoustic features, which are extracted with the frame size is 25ms and the frame shift is 10ms. The global CMVN is used to normalize the features, and 3-times downsample is applied to reduce the input frames of encoder.

AISHELL-1 has 4235 output units in total, which consist of 4230 Chinese characters and 5 extra tokens(an unknown token ($<$UNK$>$), a padding token ($<$PAD$>$), and two sentence start tokens($<$R2L$>$, $<$L2R$>$) and  one end tokens ($<$EOS$>$)).

According to~\cite{dong2018speech, mohamed2019transformers}, the transformer for end-to-end speech recognition prefers more layers of encoder than decoder. Thus we build our speech transformer baseline and STBD with 8 layers encoder and 4 layers decoder. We use $d_{m}=512, d_{f}=2048$ and the number of heads is 8. The Adam optimizer is used with warmup strategy, and we set $\beta_{1}=0.9, \beta_{2}=0.98, \epsilon=10^{-9}$. The learning rate is computed as Eq~\ref{eq:lr}:
\vspace{-0.5em}
\begin{equation}
  \text{lr} = k \cdot \min(\text{step}^{-0.5}, \text{step} \cdot \text{warmup\_steps}^{-1.5})
  \label{eq:lr}
\end{equation}
where we set $k=1.0, \text{warmup\_steps}=16000$. The feed-forward dropout is set to 0.2 to prevent overfitting. We train the ST-L2R, ST-R2L and STBD for 70 epochs with 4 GPUs. 
To improve the training efficiency, the batch size is decided dynamically by the length of audios during the training process.
Finally, we choose the best 5 epochs with lowest CER on dev set, and average them to get the final model. In inference stage, we use a narrow beam size of 2 and the length penalty is set to 0.6.

\subsection{Quantitative Results}
The overall comparison results are listed in Tab~\ref{tab:cer}. Compared with speech transformer baseline in \cite{fan2019speaker}, we implement a stronger baseline with the similar structure and denote it as \textbf{ST-L2R}.

In Tab~\ref{tab:cer}, our strong baseline ST-L2R performs better than LFMMI (chain model)~\cite{bu2017aishell} in kaldi tools with 2.2\% relative CER gain. By exploiting the bidirectional targets and bidirectional contexts, our proposed \textbf{STBD} model can improve the conventional L2R speech transformer with a margin(test CER from \textbf{7.45}\% to \textbf{7.18}\%), and achieves \textbf{5.8}\% relative CER gain compared with LFMMI. We also present the right-to-left speech transformer model(denoted as \textbf{ST-R2L}) as the reference. From the perspective of CER, there is not much difference between ST-L2R and ST-R2L. What't more, the two unidirectional decoders in STBD share their weight, thus in Tab~\ref{tab:cer}, we could observe that there are no significant difference in parameters between the STBD and the ST-R2L. We further investigate the STBD with bigger encoder-decoder structure, and refer it as \textbf{STBD-Big}, which can achieve \textbf{5.8}\% for Dev CER and \textbf{6.64}\% for Test CER. Compared with other previous works, STBD-Big achieves a large margin improvement.



\begin{figure}[t]
  \centering
  \includegraphics[width=0.7\linewidth]{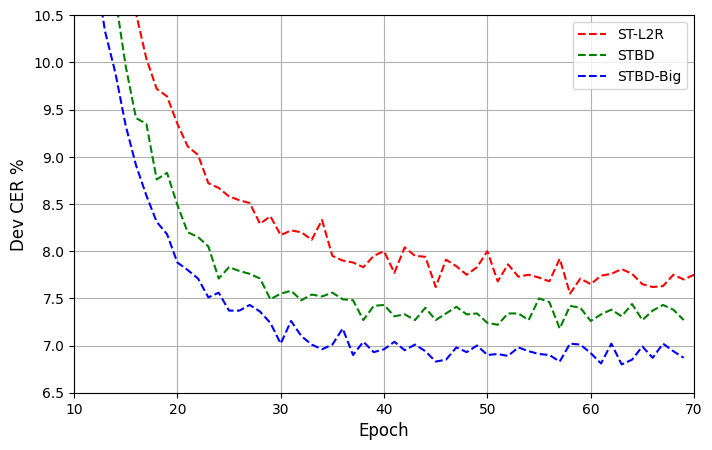}
  \vspace{-1em}
  \caption{Validation curve of baselines and our models.}
  \label{fig:devcer}
\end{figure}

We compare the training process of our STBD models with the strong baseline by plotting validation performance curve during training in Fig.\ref{fig:devcer}. It is evident that the STBD is able to improve the convergence and the final performance significantly, which indicates that the bidirectional decoder efficiently.

\begin{figure}[t]
  \centering
  \includegraphics[width=0.85\linewidth]{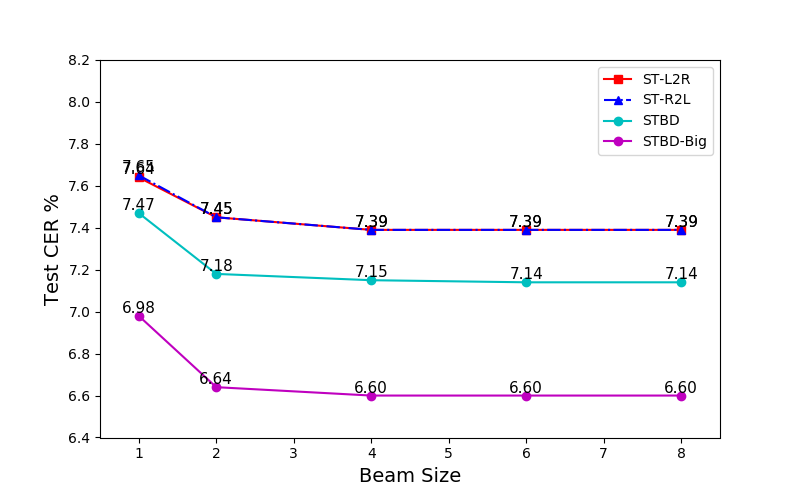}
  \vspace{-1em}
  \caption{CER with different beam size}
  \label{fig:beam}
\end{figure}

\vspace{-1em}
\subsection{Ablation Analysis}

\subsubsection{Influence of Beam Size}
We first investigate various beam size of STBD and baseline models, and plot the performance curve in Fig~\ref{fig:beam}. When the beam size is increased from 1 to 2, the improvements are significant for all models. The performance gain will be saturate when we increase the beam size further. Compared with baseline models, our STBD performs better in all different beam sizes.


\vspace{-0.5em}
\subsubsection{Effects of Bidirectional Beam Search}
We also conduct comprehensive experiments to verify the effect of bidirectional beam search method. We report the experiments results of with/without bidirectional search in Tab~\ref{tab:bs}. 
\begin{table}[th]
  \caption{CER w/o bidirectional beam search}
  \label{tab:bs}
  \centering
    \resizebox{0.3\textwidth}{!}{
  \begin{tabular}{ c c c }
    \toprule
    \textbf{Model} & 
    \textbf{Dev CER} &
    \textbf{Test CER} \\
    \midrule
    ST-L2R &  6.52\%  & 7.45\%  \\
    \textbf{STBD}   & \textbf{6.23}\% & \textbf{7.18}\% \\
    STBD-BS$_{\text{L2R}}$ & 6.39\% & 7.33\%\\
    STBD-BS$_{\text{R2L}}$ & 6.48\% & 7.39\%\\
    \bottomrule
  \end{tabular}
  }
\end{table}

The default STBD model is trained with bidirectional targets, and we conduct the bidirectional beam search during the inference process as described in Fig~\ref{fig:1b}. 
We also conduct the standard beam search on the trained STBD model to generate the prediction and denote these two methods as STBD-BS$_{\text{L2R}}$ and STBD-BS$_{\text{R2L}}$.

In Tab~\ref{tab:bs}, compared with ST-L2R, the STBD-BS$_{\text{L2R}}$ and STBD-BS$_{\text{R2L}}$ can also have a little improvement even we use the standard unidirectional beam search, which indicates optimization with the bidirectional targets is helpful for the speech recognition. We check the decoding results and find that the error accumulation exists in the unidirectional baseline, that is, the current character recognition error may affect the following characters recognition. With the error accumulation from different directions as the regularization, this problem can be alleviated in STBD, so that the STBD can gain a little improvement even with standard beam search.

Additionally, we find that some characters are easier to infer from the forward context and some characters are easier to infer from the reverse context. This should explain that the bidirectional beam search can improve the performance of the model.
In the experiment, we find that 49.4\% of sentences are decoded by backward decoder.
\vspace{-0.5em}
\subsubsection{The Visualization of Attention Alignment}
\begin{figure}[t]
  \centering
  \includegraphics[width=0.7\linewidth]{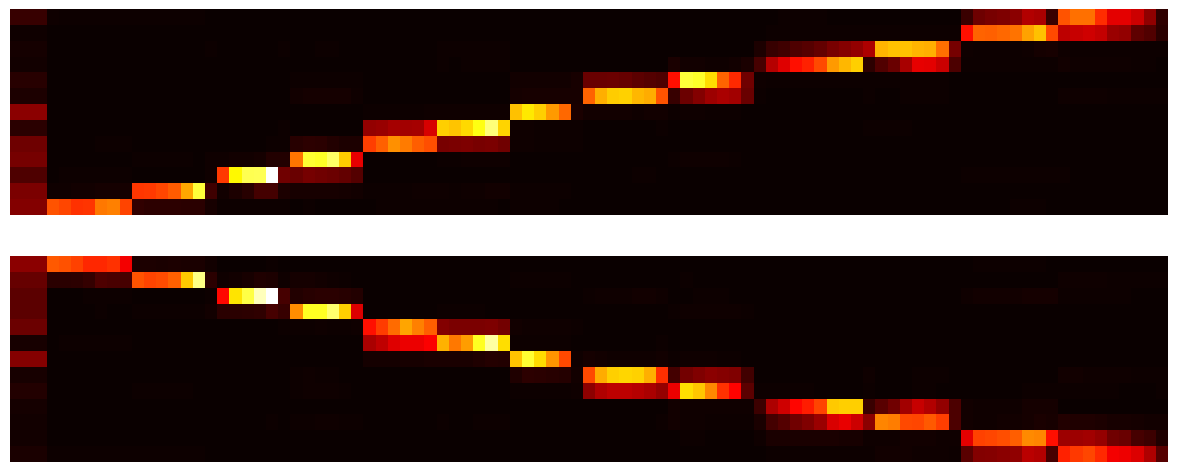}
  \caption{The attention alignment of bidirectional decoder. 
The horizontal axis represents the audio timeline, and the vertical axis represents the decoded hypothesis. The brighter color means the higher attention score. The \textbf{top} figure is the left-to-right decoder, and the \textbf{bottom} figure is the right-to-left decoder in STBD.}
  \label{fig:align}
\end{figure}

We also visualize the vanilla-attention scores between encoder and decoder in Fig~\ref{fig:align} for better understanding our proposed bidirectional decoder.
From the Fig~\ref{fig:align}, we find that the attention score of the left-to-right decoder is a diagonal line from the lower left corner to the upper right corner, and the attention score of the right-to-left decoder is a diagonal line from the lower right corner to the upper left corner. 

%
\vspace{-0.5em}
\section{Conclusions}
In this work, we have proposed a novel speech transformer with directional decoder(STBD), which can exploit both of the right-to-left contexts and left-to-right contexts. The STBD model simultaneously generates different directional targets, and utilize the bidirectional beam search synchronously, in order to update the left-to-right and right-to-left candidates. Finally, we choose the hypothesis sentences with the highest score as final prediction, which may be leaded by $\left<\text{L2R}\right>$ or $\left<\text{R2L}\right>$. Furthermore, we demonstrate the efficacy of the STBD model by extensive experiments on the AISHELL-1 dataset, which clearly show out our model has achieved competitive performance even without language model rescoring and additional data enhancement strategies.

\bibliographystyle{IEEEtran}

\bibliography{mybib}


\end{document}